\documentclass[runningheads]{llncs}
\usepackage{multirow}
\usepackage{amsmath} 
\usepackage{amssymb}
\usepackage{booktabs, pifont}
\newcommand{\cmark}{\ding{51}}  
\newcommand{\xmark}{\ding{55}}  
\usepackage{cite} 

\usepackage[hyphens]{url}
\usepackage{hyperref}

\usepackage{array}

\usepackage{booktabs}

\usepackage[T1]{fontenc}
\usepackage[dvipsnames]{xcolor}

\usepackage{mathtools}
\usepackage{graphicx, verbatim}
\usepackage{placeins}
\begin{document}
\title{Hierarchical Diffusion Framework for Pseudo-Healthy Brain MRI Inpainting with Enhanced 3D Consistency}
%
%

\author{%
  Dou~Hoon~Kwark\inst{1}\index{Kwark, Dou~Hoon} \and
  Shirui~Luo\inst{2}\index{Luo, Shirui} \and
  Xiyue~Zhu\inst{1}\index{Zhu, Xiyue} \and
  Yudu~Li\inst{1}\index{Li, Yudu} \and
  Zhi\mbox{-}Pei~Liang\inst{1}\index{Liang, Zhi‑Pei} \and
  Volodymyr~Kindratenko\inst{1,2}\index{Kindratenko, Volodymyr}
}
%
\authorrunning{Kwark et al.}
%
\institute{
The Grainger College of Engineering\\
\and
National Center for Supercomputing Applications\\
University of Illinois Urbana‑Champaign, Urbana, IL, USA\\
\email{dkwark2@illinois.edu}
}


\maketitle              
%


\begin{abstract}
Pseudo-healthy image inpainting is an essential preprocessing step for analyzing pathological brain MRI scans. Most current inpainting methods favor slice-wise 2D models for their high in‑plane fidelity, but their independence across slices produces discontinuities in the volume. Fully 3D models alleviate this issue, but their high model capacity demands extensive training data for reliable, high-fidelity synthesis---often impractical in medical settings. We address these limitations with a hierarchical diffusion framework by replacing direct 3D modeling with two perpendicular coarse-to-fine 2D stages. An axial diffusion model first yields a coarse, globally consistent inpainting; a coronal diffusion model then refines anatomical details. By combining perpendicular spatial views with adaptive resampling, our method balances data efficiency and volumetric consistency. Our experiments show our approach outperforms state-of-the-art baselines in both realism and volumetric consistency, making it a promising solution for pseudo-healthy image inpainting. Code is available at \url{https://github.com/dou0000/3dMRI-Consistent-Inpaint}.

\end{abstract}
%
%
\section{Introduction}
Brain MRI images are widely used for diagnosing and analyzing neurological conditions, but the presence of abnormalities, such as lesions and resection cavities, often disrupts automated image processing \cite{guo2019}. Common neuroimaging tools---FreeSurfer \cite{fischl2012freesurfer}, FSL \cite{Jenkinson2012}, and SPM \cite{ashburner2009computational}---are generally optimized for healthy brains, resulting in suboptimal performance when confronted with structural irregularities \cite{seiger2018cortical,pollak2025fastsurfer}. To address this, pseudo-healthy image inpainting has emerged as a preprocessing step that replaces pathological regions with synthetic healthy tissue, facilitating downstream tasks such as segmentation and registration \cite{iglesias2023synthsr,radwan2021virtual}. However, traditional approaches---like atlas-based or statistical methods---struggle to generalize across diverse patient populations and complex pathologies \cite{2013crumregistrationfail}. Recently, deep learning-based inpainting using generative models (e.g., GANs, VAEs, diffusion models) has shown promise for inpainting anatomically plausible healthy structures \cite{Liu2023Inpainting,pollak2025fastsurfer,zhu2023make,durrer2024denoising,hassanaly2024inpaintvae}. Many of these methods employ 2D slice-wise approaches, which yield high-fidelity slices and benefit from data efficiency---each 3D volume can be decomposed into multiple 2D slices \cite{2023durrerinpaintddpm,2021liuinpaint}. However, independent slice-wise processing can introduce misalignment across a stack of slices, reducing volumetric coherence. In contrast, fully 3D models capture global spatial dependencies, but their large parameter counts require extensive training data for high-fidelity synthesis \cite{zhu2025diff,2024bieder3dddpm}. In data-scarce settings, typical of medical AI, these 3D models often fail to generalize, producing low-quality synthesis.

Various strategies have emerged to balance data efficiency and volumetric consistency. For instance, Zhu et al. \cite{zhu2023make} introduced Make-A-Volume, a slice-wise latent diffusion model \cite{rombach2022ldm} for cross-modality translation. By adding volumetric layers (1D depth-wise convolutions), the model learns inter‑slice spatial relationships and thereby improves volumetric coherence. Durrer et al. \cite{durrer2024denoising} adopted this pseudo-3D approach---inspired by the Make-A-Volume method---and compared 2D, pseudo-3D (Make-A-Volume), 3D latent \cite{rombach2022ldm,Khader2022ldm} and wavelet diffusion \cite{Friedrich2024wdm} models for 3d pseudo-healthy inpainting. They found that the pseudo-3D model achieved the best overall performance. Despite its strength, this method is constrained by a fixed per-batch slice count (e.g., 16 slices in the $z$-direction\footnote[1]{\scriptsize \url{https://github.com/AliciaDurrer/DM_Inpainting}}), dictated by GPU memory. Consequently, large tumors or resection cavities that extend beyond this window provide incomplete global context, leading to sub-optimal solution, as demonstrated in Fig \ref{vis_fig}.

In this work, we introduce a hierarchical diffusion framework that enhances volumetric coherence while retaining slice-wise data efficiency by combining perpendicular spatial modeling with integrated global and local context. Specifically, our contributions are as follows: (1) We show that coupling \emph{adaptive resampling} with a two-stage perpendicular \emph{coarse-to-fine pipeline}---axial inpainting followed by coronal refinement--- enables globally consistent, high‑fidelity inpainting using only 2D diffusion models, eliminating the need for data-hungry 3D networks; (2) We introduce a \emph{\textbf{T}issue‑Aware \textbf{A}ttention \textbf{M}odule (TAM)} that implicitly guides the refinement stage to distinguish tissue types---producing anatomically coherent outputs and boosting downstream tissue-segmentation accuracy; and (3) We demonstrate on benchmark datasets that our approach outperforms state-of-the-art baselines in realism, volumetric consistency, and downstream segmentation---highlighting its potential for automated neuroimaging pipelines.

\begin{figure}
\includegraphics[width=\textwidth]{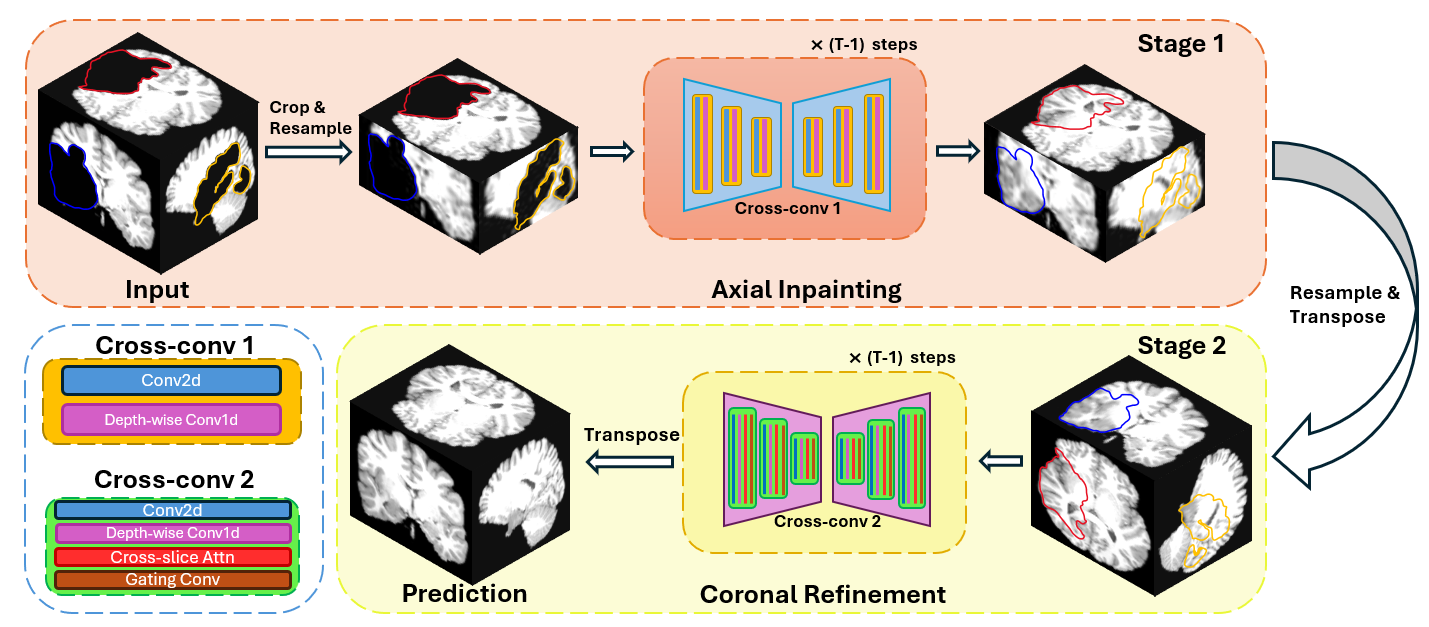}
\caption{Inference in our hierarchical framework for 3D inpainting: at each stage, the model denoises $(x_t, t)$ conditioned on the context image and mask for
$T\!-\!1$ steps, yielding $\hat x_0$ (the noise sample $x_t$ and mask are omitted from the schematic for clarity).} \label{model_fig}
\end{figure}

\section{Hierarchical Diffusion Framework for 3D Inpainting}
We propose a two-stage hierarchical framework for inpainting dynamic size of pseudo-healthy tissues in pathological brain MRI scans, preserving volumetric consistency and high-resolution details (Fig. \ref{model_fig}). The framework operates in two complementary orientations—axial and coronal—to effectively handle lesions of varying sizes and refine subtle discontinuities across the 3D volume.

\subsection{Preliminary}
Denoising Diffusion Probabilistic Models (DDPMs) \cite{ho2020denoising} have become powerful tools for image synthesis. They model a forward diffusion process that gradually corrupts a clean image $x_0$ with noise over $T$ steps and a reverse process that denoises the corrupted images. Formally, the forward step adds Gaussian noise according to 
\begin{equation}
q(x_t | x_{t-1}) = \mathcal{N}(x_t; \sqrt{1-\beta_t}x_{t-1}, \beta_t \mathbf{I}),
\end{equation}
where $\beta_t$ is a noise schedule. The model parameters $\theta$ are optimized by minimizing the mean‑squared
error between the true noise $\varepsilon$ and the network prediction
$\varepsilon_\theta$:

\begin{equation}
\mathcal{L}_{\text{MSE}}
  = \mathbb{E}_{\,t,x_0,\varepsilon_t}\!
    \bigl\|
      \varepsilon_t \;-\; \varepsilon_\theta(x_t, t)
    \bigr\|_2^2,
\end{equation}

During training for image-to-image translation, given each clean image $x_{0}$ and binary mask $m\in\{0,1\}$ ($m=1$ inside the region to fill or refine), we build a context image $x_{m}=x_{0}\odot(1-m)$ for inpainting---or $x_{m}=\mathrm{Blur}(x_{0})\odot m + x_{0}\odot(1-m)$ for refinement---then add Gaussian noise to obtain $x_{t}\sim q(x_{t}\!\mid x_{0},t)$, and feed the tuple $(x_{t},x_{m},m,t)$ to the diffusion denoising model (here, $\odot$ denotes the Hadamard multiplication); the model is thus trained to remove the noise and reconstruct $x_{0}$ only within $m$ while leaving the unmasked context unchanged.

\subsection{Axial inpainting}
We begin our framework with a 2D inpainting model in the \emph{axial} view, producing fine in-plane details and coarse---but globally consistent---out-of-plane structure.

\textbf{Depth-wise 1D convolution.}
Conventional 2D DDPMs achieve excellent in-plane fidelity but have limited awareness of inter-slice dependencies, which often results in discontinuities across adjacent slices. To capture inter-slice dependencies without relying on costly full 3D kernels, following \cite{zhu2023make,durrer2024denoising}, we append 1D convolutional filters to 2D convolutional blocks, which can be viewed as a variant of separable convolution  \cite{chollet2017xception}. Specifically, cross-slice interactions are captured by reshaping each feature map from $(b, c, h, w)$ to $(h \times w, c, b)$, treating the spatial dimensions as an extended batch and applying depth-wise 1D convolution. The 1D filter is Dirac-initialized (e.g., identity-like) so that early in training the network behaves as a purely 2D model, gradually learning cross-slice mixing over time.

\textbf{Adaptive resampling.} Although depth-wise 1D convolutions enlarge the out-of-plane receptive field, the context they capture in the superior-inferior ($z$) axis remains inherently constrained by the number of slices that fit into GPU memory. This limitation often hampers global reasoning---especially for large lesions spanning many slices---and can cause severe artifacts in the non-linear ill-posed settings like inpainting (Fig. \ref{vis_fig}). To provide full volumetric context without exceeding memory limits, we resample exclusively in the $z$-dimension: the volume is first cropped to the bounds of the masked region in $z$ axis, then uniformly rescaled such that the volume depth does not exceed a hardware ceiling $Z_{max}$. Because the in-plane details are left unchanged, this step simply increases the out-of-plane slice spacing---typically to 2-5 mm in datasets such as HCP and BraTS. This adaptive resampling can be viewed as a variable‑slice‑spacing augmentation \cite{billot2023synthseg} that trains the model to stay robust to out-of-plane resolution shifts while still producing high‑quality outputs. Once axial inpainting is completed on the resampled volume, we restore the original slice count via cubic interpolation and pass the result to the coronal refinement stage to recover fine $z$-axis details.

\subsection{Coronal Refinement}
\label{coronalrefinement}

Although axial inpainting produces high in-plane detail, the interpolation required to reverse the resampling inevitably blurs structures along the superior-inferior axis. We therefore apply a second 2D diffusion model in the \emph{coronal} view to restore the missing detail.

\textbf{Perpendicular Modeling and Coarse-to-Fine.} 
Lee et al. \cite{lee2023perpendicular} showed that running two perpendicular 2D diffusion models in parallel can approximate volumetric generation. Inspired by this, we adopt a \emph{sequential} coarse-to-fine strategy: an axial model produces a globally coherent but slightly over-smoothed volume, and a coronal model subsequently predicts only the residual necessary to restore anterior-posterior details. By formulating the second stage as a residual task \cite{yu2019freeform,tian2021coarsetofine}, we reduce the objective from full volumetric generation to refining only the remaining local, high-frequency discrepancies, while depth-wise 1D convolutions preserve cross-slice continuity. Together, these design choices enable the refinement stage to recover fine anatomical structures---such as key distinctions among white matter (WM), gray matter (GM), and cerebrospinal fluid (CSF)---that were ignored during the first stage.

\textbf{Tissue-Aware Attention Module.} Depth-wise 1D convolutions provide inter-slice context, yet their shared weights blend WM, GM, and CSF features indiscriminately, failing to capture tissue boundaries effectively and weakening tissue‑specific contrast. To provide content-adaptive mixing, we introduce a \textbf{T}issue-Aware \textbf{A}ttention \textbf{M}odule (TAM) that combines token-based Cross-Slice Attention with Gated Convolution. Specifically, \textit{(i)} we spatially average-pool each slice's feature map to a token $t_y \!\in\! \mathbb{R}^{B\times C}$, \textit{(ii)} perform multi-head self-attention \cite{vaswani2017attention} over these tokens to capture slice-level dependencies, and \textit{(iii)} broadcast the attended token back to every spatial location in the slice. We then employ a free-form gated convolution \cite{yu2019freeform}, initialized so that $\sigma(\text{bias})\approx 1$, allowing the network to learn to adaptively emphasize or suppress new features during optimization. As shown in Tab. \ref{tab:abl}, TAM improves tissue differentiation among WM/GM/CSF in the generated volumes. Combined with our hierarchical coarse-to-fine pipeline, it enables pseudo-healthy inpainting that is anatomically coherent at both global and voxel levels.

\section{Experiments and Results}
\subsection{Dataset}
\label{dataset}
\textbf{BraTS 2023 Local Synthesis}~\cite{kofler2023brain}. 
The training set comprises 1,251 bias-corrected T1 volumes, resampled to $1\text{mm}^3$ and cropped to the spatial dimension of $240\!\times\!240\!\times\!155$. Masks for pseudo-healthy inpainting are provided within the healthy tissue, according to \cite{kofler2023brain}; they are relatively small because tumors in many BraTS volumes occupy large portions of the brain. As official validation/test labels are not publicly available, we only used the training set, randomly reserving 251 subjects for testing and the remaining 1,000 for training. 

\noindent\textbf{Human Connectome Project (HCP).} The HCP T1 subset \cite{2021elamHCP} contains 1,061 healthy brains at $0.7\text{mm}^3$ on the same dimension of $240\!\times\!240\!\times\!155$, after bias field correction and spatial normalization. To obtain inpainting masks for inpainting assessment, we transplanted BraTS tumor masks (note that they are distinct from the small healthy masks used for BraTS inpainting) into these tumor-free volumes, following the same protocol as \cite{kofler2023brain} to preserve metholdological consistency and ensure the validity of the pseudo-healthy task. In BraTS, existing tumor regions limit inpainting to small masks for evaluation, whereas the tumor-free HCP brains allows us to utilize lesions of arbitrary size and location, enabling richer evaluation of inpainting performance. We used 213 volumes for testing and the remaining 848 for training.

\subsection{Implementation Details}

All volumes are zero-padded to ${256\times256\times160}$ for both BraTS and HCP datasets. Our diffusion framework was built on \cite{dhariwal2021diffusion} with a linear noise scheduling spanning ${1\times10^{-4}}$ to ${2 \times10^{-2}}$ over ${T=1000}$ steps, using concatenation-based conditioning \cite{durrer2024denoising}.  Both stages are trained for $3.25\times10^{5}$ iterations on an NVIDIA A100 GPU, using the Adam optimizer with a learning rate of $10^{-4}$; batch sizes are 24 (e.g., $Z_{max}$ = 24) and 16 (stage 2) for each stage 1 and 2.

\begin{figure}[t!]
\includegraphics[width=\textwidth]{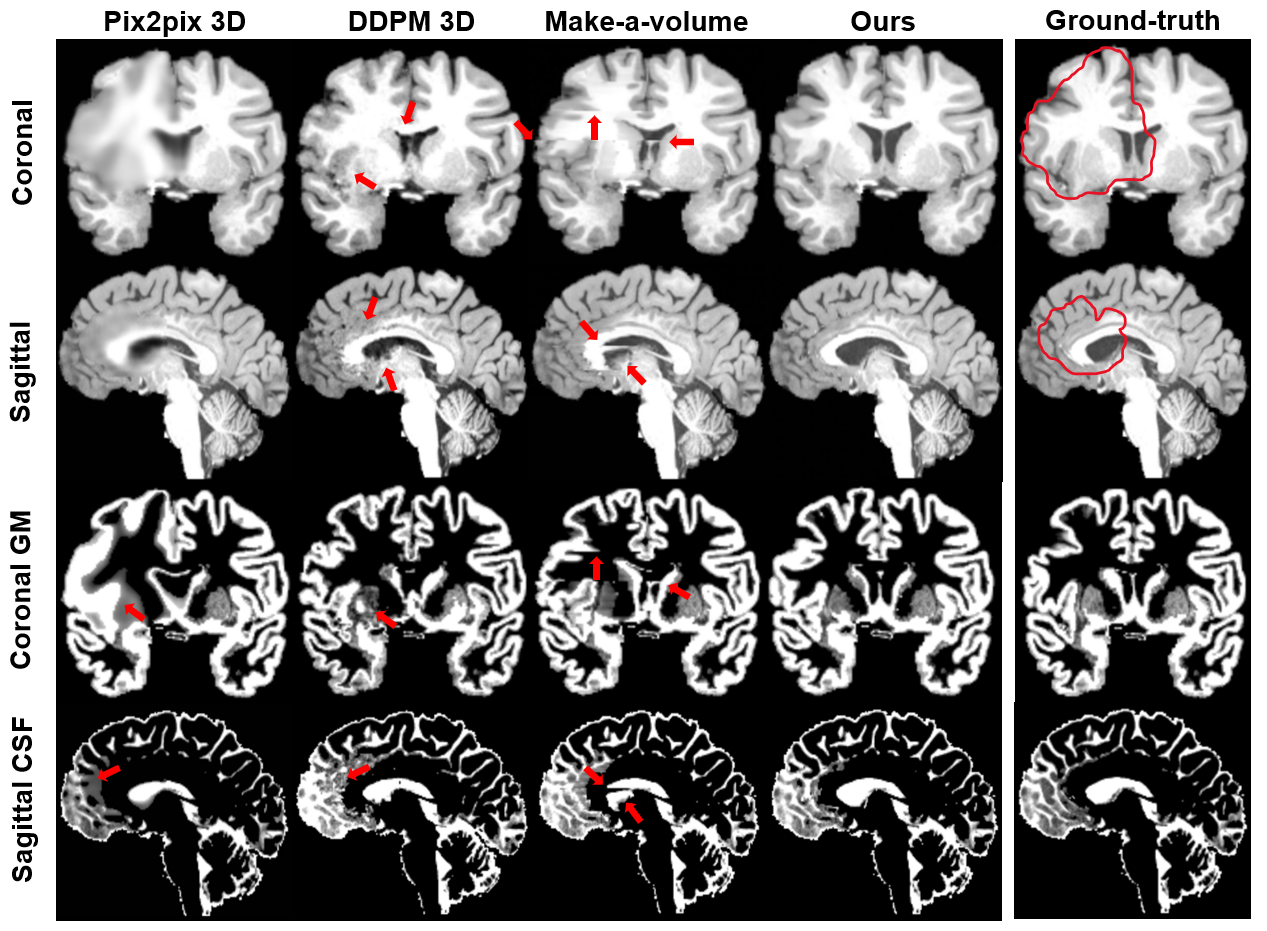}
\caption{\textbf{Qualitative Comparison.} Qualitative analysis on HCP dataset for generation and segmentation using FAST. The inpainting region is indicated by red boundary.} \label{vis_fig}
\end{figure}

\subsection{Performance Evaluation}
To evaluate our proposed framework, we compared it against four baseline models. We selected 3D Autoencoder \cite{hinton2006reducing} and Pix2Pix \cite{isola2017img2img} for their widely recognized success in generative tasks, and a memory-efficient 3D DDPM \cite{2024bieder3dddpm} for comparison with direct volumetric diffusion-based generation. Finally, we included Make-A-Volume \cite{zhu2023make} as prior work shows that it outperforms other recent diffusion variants---most notably latent-diffusion \cite{rombach2022ldm,Khader2022ldm} and wavelet-diffusion \cite{Friedrich2024wdm}---in volumetric inpainting \cite{durrer2024denoising}. Image-level fidelity was measured within the lesion mask using structural similarity (SSIM), peak signal-to-noise ratio (PSNR), and mean squared error (MSE). Anatomical plausibility was assessed by running FAST \cite{Zhang2001fast} and FastSurfer \cite{Henschel2020fastsurfer} on the inpainted HCP volumes to obtain CSF, GM, and WM segmentations; We then computed Dice scores within the masked region using healthy ground-truth data. We focused segmentation evaluation on the HCP dataset because BraTS volumes contain tumors, leaving only small regions of healthy tissue for inpainting, which makes CSF/GM/WM segmentation evaluation less meaningful.

\begin{table}[h]
\centering
\caption{Comparison of performance metrics in masked areas for both datasets.}
\label{tab:tab1}
\begin{tabular}{llccc}
\toprule
\textbf{Dataset} & \textbf{Model} & \textbf{SSIM $\uparrow$} & \textbf{MSE $\downarrow$} & \textbf{PSNR $\uparrow$} \\
\midrule
\multirow{5}{*}{\textbf{HCP}} 
    & Autoencoder 3D       & $0.510\pm0.053$  & $0.024\pm0.008$  & $16.385\pm1.283$ \\
    & Pix2pix 3D  & $0.554\pm0.055$  & $0.019\pm0.006$  & $17.333\pm1.239$ \\
    & DDPM 3D     & $0.471\pm0.059$              & $0.034\pm0.009$              & $14.828\pm1.160$ \\
    & Make-A-Volume  & $0.504\pm0.074$  & $0.027\pm0.011$  & $15.965\pm1.536$ \\
    & \textbf{Ours}   & \textbf{0.598$\pm$0.065} & \textbf{0.017$\pm$0.006} & \textbf{17.845$\pm$1.381} \\
\midrule
\multirow{5}{*}{\textbf{BraTS}} 
    & Autoencoder 3D       & $0.729\pm0.147$  & $0.020\pm0.013$  & $17.674\pm2.585$ \\
    & Pix2pix 3D  & $0.758\pm0.136$  & $0.015\pm0.011$  & $18.834\pm2.628$ \\
    & DDPM 3D     & $0.674\pm0.178$           & $0.045\pm0.044$              & $15.866\pm4.794$ \\
    & Make-A-Volume  & $0.793\pm0.136$  & $0.009\pm0.007$  & $21.912\pm4.318$ \\
    & \textbf{Ours}   & \textbf{0.805$\pm$0.126}  & \textbf{0.008$\pm$0.006}   & \textbf{22.039 $\pm$3.789} \\
\bottomrule
\end{tabular}
\end{table}

\noindent\textbf{Qualitative analysis} As shown in Fig.\ref{vis_fig}, our method produces more realistic and anatomically consistent inpainted regions than the baseline models. Compared to Make-A-Volume---which shows visible stripe artifacts (red arrows)---our approach preserves global structure and maintains finer anatomical details, particularly around cortical surfaces and in the ventricular CSF region. In addition, while the 3D DDPM produces coherent volumetric shapes overall, it still exhibits diminished detail and occasional over-smoothing around critical tissue boundaries. To further demonstrate the anatomical plausibility of each method, Fig.\ref{vis_fig} provides tissue segmentation of the inpainted volumes. Our method yields more accurate tissue boundaries, reflecting improved tissue synthesis compared to Make-A-Volume and 3D DDPM. Specifically, Make-A-Volume’s limited volumetric understanding along the superior–inferior axis induces stripe artifacts and misalignment, which degrade segmentation accuracy, and the 3D DDPM---though globally consistent---may still miss the subtle structural cues required for fine-grained segmentation. In contrast, our two-stage strategy leverages stronger global structural and in-plane local detail priors, resulting in high-fidelity volumetric inpainting as well as robust local anatomical detail, thus offering both overall image quality and downstream segmentation tasks.

\begin{table}[h!]
\centering
\caption{Dice scores for WM, CSF, and GM segmentations in masked areas on the HCP dataset, evaluated using FSL FAST and FastSurfer.}
\label{tab:segmentation_swapped}
\begin{tabular}{llccc}
\toprule
\textbf{Tools} & \textbf{Models} & \textbf{WM} & \textbf{CSF} & \textbf{GM} \\
\midrule
\multirow{5}{*}{\textbf{FAST}} 
& Autoencoder 3D  & $0.755\pm0.066$ & $0.444\pm0.079$ & $0.656\pm0.055$ \\
 & Pix2pix 3D      & $0.787\pm0.053$ & $0.538\pm0.060$ & $0.689\pm0.053$ \\
 & DDPM 3D         &    $0.776\pm0.066$     & 
 $0.475\pm0.070$      & $0.656\pm0.046$       \\
 & Make-A-Volume   & $0.780\pm0.053$ & $0.537\pm0.066$ & $0.692\pm0.043$ \\
 & Ours            & \textbf{0.827$\pm$0.047} & \textbf{0.627$\pm$0.059} & \textbf{0.742$\pm$0.038} \\
\midrule
\multirow{5}{*}{\textbf{FastSurfer}} 
& Autoencoder 3D  & ${0.742\pm0.069}$       & ${0.412\pm0.178}$       & ${0.531\pm0.096}$       \\
 & Pix2pix 3D      & ${0.773\pm0.056}$       & ${0.617\pm0.169}$       & ${0.587\pm0.094}$       \\
 
 & DDPM 3D         & ${0.769\pm0.064}$       & ${0.640\pm0.140}$       & ${0.653\pm0.066}$       \\
 & Make-A-Volume   & ${0.773\pm0.061}$       & ${0.676\pm0.136}$       & ${0.672\pm0.055}$       \\
 & Ours            & \textbf{0.829$\pm$0.051}       & \textbf{0.780$\pm$0.103}       & \textbf{0.737$\pm$0.055}       \\
\bottomrule
\end{tabular}
\end{table}

\noindent\textbf{Quantitative analysis.} We evaluated inpainting fidelity on BraTS and HCP using SSIM, PSNR, and MSE (Table~\ref{tab:tab1}), observing that our model consistently outperforms all baseline approaches. On BraTS volumes, where most inpainting masks are relatively small, we noted only modest gains compared to Make-A-Volume. However, as the inpainting regions grew larger, our method’s advantage became more pronounced---indicating that it excels when broader structural context is required. We then assessed downstream tissue segmentation performance on the HCP dataset (Table~\ref{tab:segmentation_swapped}) using FSL FAST and FastSurfer. Our method again surpassed other methods, achieving consistently higher Dice scores for WM, CSF, and GM. Interestingly, although the DDPM 3D preserves global volumetric coherence, Make-A-Volume’s 2D architecture can yield better local fidelity, improving segmentation relative to the DDPM 3D---demonstrating that direct 3D training can be sub-optimal in data‑scarce settings. Notably, Pix2Pix attains SSIM/PSNR values comparable to ours on the HCP dataset because these metrics favor smoothed, mean-like outputs \cite{dohmen2025metrics}; its blurred predictions, however lack fine anatomical detail, resulting in noticeably lower Dice scores for CSF/GM segmentation. Overall, our two-stage approach offers a balanced synergy of global volumetric consistency and fine-grained tissue synthesis, thus leading to superior structural synthesis and higher segmentation accuracy across all tissue types.

\subsection{Ablation Analysis}
To quantify the impact of our tissue-aware module, TAM, at the refinement stage, we performed an ablation study on the inpainted HCP volumes, evaluating downstream tissue segmentation---our primary proxy for the inpainting model's practical utility. Table~\ref{tab:abl} shows that incorporating TAM improves Dice scores for CSF, GM, and WM---indicating enhanced tissue differentiation. These results suggest TAM provides a stronger anatomical prior, enabling more coherent volumes and benefiting downstream segmentation tasks.

\begin{table}[h!]
  \centering
  \caption{Dice scores within masks on inpainted HCP regions. Parentheses show the relative reduction in residual segmentation error achieved by TAM, computed as $1 - (1-\text{Dice}_{\text{TAM}})/
                (1-\text{Dice}_{\text{noTAM}})$.}
  \label{tab:abl}
  \footnotesize
  \resizebox{\linewidth}{!}{%
    \begin{tabular}{c ccc ccc}
      \toprule
      \multirow{2}{*}{\textbf{TAM}} &
        \multicolumn{3}{c}{\textbf{FAST}} &
        \multicolumn{3}{c}{\textbf{FastSurfer}} \\
      \cmidrule(lr){2-4}\cmidrule(lr){5-7}
        & \textbf{CSF} & \textbf{GM} & \textbf{WM} &
          \textbf{CSF} & \textbf{GM} & \textbf{WM} \\ \midrule
      \xmark & 0.615 & 0.734 & 0.824
             & 0.770 & 0.730 & 0.827 \\[2pt]
      \cmark & 0.627\;(\textbf{+3.1\%}) & 0.742\;(\textbf{+3.0\%}) & 0.827\;(\textbf{+1.7\%})
             & 0.780\;(\textbf{+4.4\%}) & 0.737\;(\textbf{+2.6\%}) & 0.829\;(\textbf{+1.2\%}) \\
      \bottomrule
    \end{tabular}%
  }
\end{table}

\section{Discussion and Conclusions}
We present a hierarchical coarse‑to‑fine diffusion framework for pseudo‑healthy brain MRI inpainting. Two perpendicular 2D stages with adaptive slice resampling replace data‑hungry 3D networks while retaining full volumetric context, and our Tissue‑Aware Attention Module sharpens anatomical details. The approach surpasses state‑of‑the‑art methods on BraTS and HCP datasets, delivering more realistic, anatomically coherent, and volumetrically consistent volumes---suggesting a promising path for robust 3D inpainting in medical imaging. Yet several challenges remain for future work. First, inference with the current 2D DDPM framework is slower than direct 3D models, and although lighter, still requires substantial GPU memory. Exploring latent-diffusion frameworks \cite{rombach2022ldm} could therefore be a promising alternative, simultaneously lowering the memory ceiling and accelerating generation. In addition, the proposed framework’s resilience to diverse and complex pathologies has yet to be rigorously assessed; systematic, multi-site evaluations on heterogeneous datasets can help establish its real-world generalizability.

\begin{credits}
\subsubsection{\ackname} This work used the Delta system at the National Center for Supercomputing Applications through allocation CIS240171 from the Advanced Cyberinfrastructure Coordination Ecosystem: Services \& Support (ACCESS) program, which is supported by National Science Foundation grants \#2138259, \#2138286, \#2138307, \#2137603, and \#2138296. This work has also been partially funded by the Jump ARCHES endowment through the Health Care Engineering Systems Center at Illinois and the OSF Foundation.

\subsubsection{\discintname}
The authors declare no competing interests.
\end{credits}

%
%
%
%

\end{document}